# Authority Signals in AI Cited Health Sources: A Framework for Evaluating Source Credibility in ChatGPT Responses

Erin Jacques[1], Erela Datuowei[2], Vincent Jones II[1], Corey Basch[3], Celeta Vanderpool[2], Nkechi Udeozo[4], Griselda Chapa[1]

[1]Department of Health and Human Performance, York College, The City University of New York, 94-20 Guy R. Brewer Blvd. Jamaica, NY 11451

[2]Department of Health Studies & Applied Educational Psychology, Teachers College, Columbia University, 525 West 120th Street, New York, NY 10027

[3]Department of Public Health, William Paterson University, 300 Pompton Road Wayne, NJ, 07470

[4]Department of Health Policy and Management, CUNY School of Public Health, 55 W 125th St, New York, NY 10027

**Corresponding Author:** ejacques1@york.cuny.edu

**Abstract**

Health information seeking has fundamentally changed since the onset of Large Language Models (LLM), with nearly one third of ChatGPT's 800 million users asking health questions weekly. Understanding the sources of those AI generated responses is vital, as health organizations and providers are also investing in digital strategies to organically improve their ranking, reach and visibility in LLM systems like ChatGPT. As AI search optimization strategies are gaining maturity, this study introduces an Authority Signals Framework, organized in four domains that reflect key components to health information seeking, starting with "Who wrote it ?" (Author Credentials), followed by "Who published it?" (Institutional Affiliation), "How was it vetted?" (Quality Assurance), and "How does AI find it?" (Digital Authority). This descriptive cross-sectional study randomly selected 100 questions from HealthSearchQA which contains 3,173 consumer health questions curated by Google Research from publicly available search engine suggestions. Those questions were entered into ChatGPT 5.2 Pro to record and code the cited sources through the lens of the Authority Signals Framework's four domains. Descriptive statistics were calculated for all cited sources (n=615), and cross tabulations were conducted to examine distinction among organization types. Over 75% of the sources cited in ChatGPT's health generated responses were from established institutional sources, such as Mayo Clinic, Cleveland Clinic, Wikipedia, National Health Service, PubMed with the remaining citations sourced from alternative health information sources that lacked established institutional backing.

**1. Introduction**

The adoption rate of AI for health guidance is on the rise with nearly one third of ChatGPT's 800 million users asking health questions weekly (OpenAI, 2025), and likely making medical decisions based on those answers. Surveys report that as high as 78% of users acknowledge using ChatGPT to self diagnosis (Shahsavar & Choudhury, 2023), and healthcare professionals show similar adoption rates. Nearly half of nurses report using AI weekly (American Medical Association, 2024) for clinical decision making (Iqbal et al., 2025), and physician use doubled from 38% to 66% in just one year (American Medical Association, 2024). With this widespread use that impacts medical outcomes, a critical question emerges: which health information sources are these systems actually citing?

It's clear that the landscape of health information seeking has changed, and much of which is attributed to OpenAI's launch of ChatGPT's in 2022, which was further enhanced after the introduction of real-time web search capabilities in 2023 (Search Engine Journal, 2025). The ease of accessing health information from personal devices coupled with a trust in physicians and hospitals having dropped dramatically post COVID (Perlis et al., 2024), has driven individuals toward non-traditional health information sources (Casselman-Hontalas et al., 2024). The shift toward consumer seeking health information using AI powered systems introduces new challenges around the credibility of those sources.

ChatGPT demonstrates capability in performing diagnoses and recommendations, in some cases outperforming doctors who have AI support (CADTH, 2025). However, there are growing ethical concerns over the validity and reliability of medical information from ChatGPT including issues of fairness, bias, transparency, and privacy (Haltaufderheide & Ranisch, 2024). Both medical professionals and the general public cite instances of misleading information and hallucinations from AI chatbots (Tangsrivimol et al., 2025).

There are numerous credibility signals that exist to help identify trustworthy health information (Fahy et al., 2014). The established markers that suggest content is factual include professional credentials (MD or PhD), institutional affiliations with government agencies or academic medical centers, peer reviewed editorial processes, and evidence-based citations (MedlinePlus, 2024). However, discerning quality health information requires health literacy, which 88% of US adults lack at a proficient level (Kutner et al., 2006). This leaves many consumers unable to recognize established credibility signals (Sun et al., 2019; Diviani et al., 2015).

The shift from web-based searching to AI systems that provide instant responses has generated the critical need for systematic quality monitoring of health information. Recent studies show that between 50% to 90% of LLM-generated health responses are not fully substantiated by their cited sources (Wu et al., 2025). This raises concerns about the efficacy of traditional credibility markers for guiding source selection in AI systems, particularly as both legitimate health organizations and alternative information providers continue to compete for visibility within AI systems with limited transparency and no standardized ranking criteria (Chen et al., 2024).

Commercial incentives to increase source visibility in AI-generated responses can amplify sources regardless of medical credibility (Chen et al., 2024). This, coupled with the lack of citation transparency in current LLM systems heightens the potential for response manipulation as optimization practices mature (Gao et al., 2023). Considering the potential for health misinformation to lead to consequences such as delayed care or adoption of ineffective treatment, source vetting and ranking becomes all the more important (Memon et al., 2022; Tong et al., 2023).

Understanding which sources AI systems actually cite, and the characteristics of those sources is necessary in evaluating whether these systems prioritize established quality markers or to what extent they are vulnerable to manipulation. Authority signals such as author credentials, institutional affiliations, and editorial review processes have long been established as foundational indicators of health information quality (Silberg et al., 1997). Identifying which authority signals are actually present in AI cited health sources is the necessary first step to understanding whether these systems prioritize established markers of quality or other criteria.

**1.1 Research Gap**

Recent research has examined ChatGPT's performance in healthcare contexts through various lenses. Systematic reviews have evaluated the accuracy of ChatGPT generated medical content, finding overall accuracy rates of 56% (Cai et al., 2024) and found accuracy in response varying based on domain (Ge et al., 2025). Studies have also examined citation reliability. Hosseini et al. (2023) found that 47% of ChatGPT generated citations were false and 93% were inaccurate.

Prior studies have focused primarily on answer accuracy and citation reliability rather than the systematic evaluation of source credibility overall. Authority signals such as author credentials, institutional affiliations, and editorial review processes have long been recognized as established indicators of health information quality (Silberg et al., 1997), yet no framework exists for systematically characterizing these signals across AI-cited health sources.

The terminology around optimization to increase source visibility in LLMs remains actively contested. Practitioners use terms including Answer Engine Optimization (AEO), Generative Engine Optimization (GEO), Generative Search Optimization (GSO), and AI Answer Engine Optimization (AI AEO), defining them with inconsistent and various nuances, and sometimes using them interchangeably. Despite the lack of industry consistency in the nuances, they describe strategies for improving visibility in large language model responses (Digiday, 2025; Backlinko, 2025). Though standardized nomenclature for these optimization techniques remains contested, the entrepreneurial landscape is already iterating methods to increase their representation in AI citation patterns.

Understanding the characteristics of the sources most likely to be cited by AI systems is necessary for evaluating whether and how vulnerable to manipulation these systems are. This study seeks to establish baseline authority signal patterns during this early experimental phase, before optimization strategies become widespread. Understanding which sources AI systems prioritize requires asking four critical questions:

1. Who wrote it? (Author Credentials)
2. Who published it? (Institutional Affiliation)
3. How was it vetted? (Quality Assurance)
4. How does AI find it? (Digital Authority)

This study introduces the Authority Signals Framework, 11 systematically coded authority signals across these four domains to characterize credibility markers present in AI cited health sources (see Figure 1).

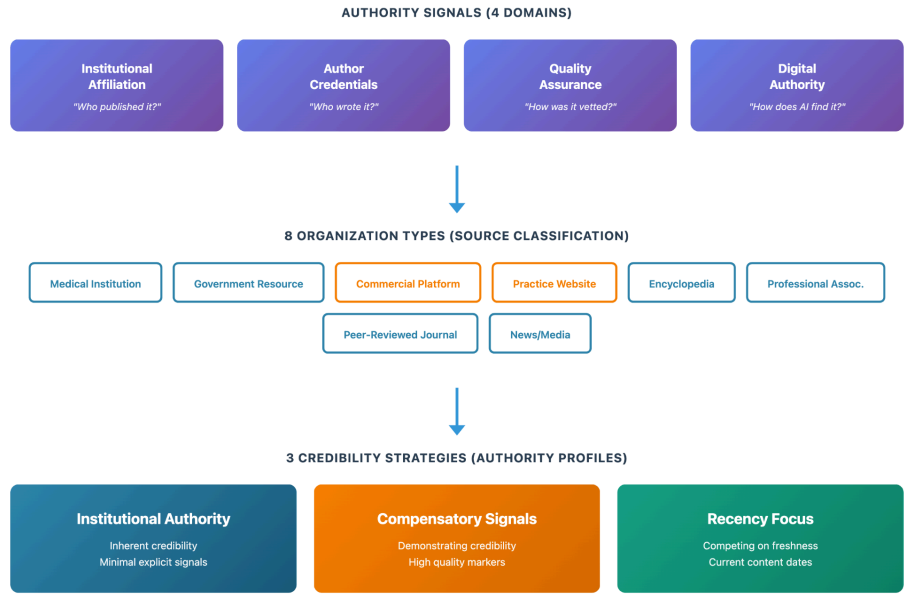

**Figure 1:** Authority Signals Framework. Conceptual model showing the relationship between authority signal domains, organization types, and credibility strategies. Four domains of authority signals (Institutional Affiliation, Author Credentials, Quality Assurance, Digital Authority) were used to classify 615 health sources into eight organization types.

The Authority Signals Framework operates through the coding of observable source characteristics across the following four domains, each addressing a critical credibility question:

**Author Credentials** answers "Who wrote it?" by capturing individual-level expertise through professional degrees, certifications, and medical specialization. Variables range from no author attribution to visible medical credentials with specialty identification.

**Institutional Affiliation** answers "Who published it?" by categorizing sources by organization type, highlighting that different organizational contexts (government agencies, medical institutions, academic institutions, commercial platforms, professional practices) carry varying levels of inherent institutional credibility in health information contexts.

**Quality Assurance** answers "How was it vetted?" by documenting editorial processes and evidence standards through presence of medical review statements, content recency, and citation of supporting references.

**Digital Authority** answers "How does AI find it?" by quantifying web-based credibility signals including Moz Page Authority, Domain Authority, spam scores, schema markup implementation, and content length.

Each variable was operationalized with explicit coding criteria to ensure reproducibility (available via Zenodo: https://doi.org/10.5281/zenodo.18287499). This approach acknowledges the intersecting signals that inform digital health information source credibility.

By applying this framework to sources cited by ChatGPT in response to 100 validated consumer health questions from the HealthSearchQA dataset, this study establishes whether AI systems prioritize established quality markers or alternative selection criteria. This knowledge is essential for understanding AI's role in the dissemination of health information.

## 2. Methods

### 2.1 Data Source and Sample

This descriptive cross-sectional study used data from HealthSearchQA which contains 3,173 consumer health questions curated by Google Research from publicly available search engine suggestions. These questions reflect common queries associated with medical search terms. The questions were randomized and the first 100 were selected for inclusion.

One coder manually entered each of the 100 health questions into ChatGPT 5.2 Pro using a newly created ChatGPT account and email to prevent prior conversation influence. All of the data was collected on January 11, 2026 with each question entered into a separate chat session to avoid subsequent responses being influenced by previous queries. To ensure sources were provided for the responses, each question was followed by the standardized prompt: "Please include sources with links in your response." Each question generated 4+ sources, resulting in 618 initial responses with each row in the Google Sheets representing one unique source. Three responses were removed due to inoperable sites, 403, and 404 errors resulting in a final sample of 615. A second coder was used to verify a subset of the chats to ensure all sources were accurately transferred to the spreadsheet.

## 2.2 Variables and Measures

The Authority signals across the four domains of Digital Authority, Quality Assurance, Institutional Affiliation, and Author Credential are individually categorized:

1. Institutional Affiliation Domain references organization type which has the following eight categories coded from 1-8: 1=Medical Institution, 2=Government Resource, 3=Commercial Health Info, 4=Professional/Practice Website, 5=Encyclopedia, 6=Professional Association, 7=Peer-Reviewed Journal, and 8=News/Media
2. Author Credentials Domain reflects an author's attribution level coded as 0=none, 1=name only, 2=name with credentials. These two author variables [Visible author attribution (yes/no); Author credentials (yes/no) ] were later collated into a combined Author attribution level
3. Quality Assurance Domain includes the four categories of references cited (yes/no); medical review statement visible (yes/no); content recency (0=before 2020 or not listed, 1=2020-2023, 2=2024-2026). Originality scores to determine whether web based content was AI or human generated scores ranged 0-100.
4. Digital Authority Domain containing six categories of page authority, domain authority, and spam score on a 0-100 scales (higher = more spam signals); content length represented as (0=brief, 1=moderate, 2=comprehensive), schema markup (yes/no) and coded as schema type (0=none, 1=microdata, 2=JSON-LD)

Complete operational definitions available via Zenodo [https://doi.org/10.5281/zenodo.18287499]

### 2.3 Measurement Procedures

**Automated Coding Pipeline**: Given the scale of the dataset (615 sources) and the need for consistent application of operationally defined authority signals, an automated coding pipeline was developed. Health website content was systematically collected using Apify for web scraping. Researcher EJ developed a codebook with the help of AI to determine clear operation definitions to serve as criteria when using Claude API to examine the categories and code details from each website. The pipeline was deployed as a serverless function on Vercel to enable batch processing of URLs. Results were stored in Supabase (PostgreSQL database) and exported to Google Sheets for validation. Any missing data fields were reviewed by both coders and entered manually, which occurred in places where the automated pipeline was unable to extract the content from the URLs. A subsequent quality control step involved JavaScript browser extension code that applied the same operational definitions to validate the coded data and flag any missing data fields. This successfully coded the following categories: Source Name, Organization Name, Organization Type, Institutional Affiliation, Author Name, Author Credentials, Temporal Date, Content Length Category, Medical Review Stated, References Cited, Number of References, Has Schema, Schema Type

Complete organizational classification mappings are available via Zenodo (https://doi.org/10.5281/zenodo.18287499)

**Content recency coding**: Content recency was determined through frequency analysis in Jamovi v2.3, revealing natural date clustering. The temporal dates were extracted from the automated pipeline and then coded in a contingency recency column following this coding convention: 0 =before 2020 or No Date Listed; 1 = 2020-2023 and 2 = 2024-2026.

**API and Bulk Coding**: Page Authority, Domain Authority Score (DA score), and Spam Score (0-100; higher = more spam signals) were collected via Moz API. Content originality was assessed using Originality.ai,

which successfully classified approximately 50% of sources (n=~309). The remaining sources showed errors in the ability to analyze for originality.

**Manual Classification**: During automated data collection, organization name and institutional affiliation were extracted for each source, yielding 248 unique organizations. Organization type was classified by reviewing organization name and institutional affiliation to determine the best category. When classification was ambiguous, a DA score was used as a third indicator to confirm organization type since higher DA scores are typically aligned with larger scale operations. In those cases, an additional review of organizational websites was conducted. Between Commercial Health Info and Professional/Practice Website, Domain Authority served as an additional classifier with DA ≤ 45 indicating Professional/Practice and DA > 45 indicating Commercial (Figure 3). An 8 category classification system was created and coded from 1-8: (1) Medical Institution, (2) Government Resource, (3) Commercial Health Info, (4) Professional/Practice Website, (5) Encyclopedia, (6) Professional Association, (7) Peer-Reviewed Journal, and (8) News/Media. Complete organization type definitions and source listings are provided in **Supplementary Appendix A**.

## 3. Analysis

Descriptive statistics were calculated for all cited sources (n=615), including frequencies, percentages, medians, and percentiles. Cross tabulations were also conducted to examine the following authority signals across the eight organization types: medical review, references cited, schema markup, content length, content recency. Software used included Google sheets for data preparation and Jamovi for statistical analysis.

## 4. Results

Citations were dominated by sources with inherent institutional authority (75.7%) including medical institutions, government resources, professional associations, peer-reviewed journals, encyclopedias, news, and media. The remaining 24.3% accounted for commercial health information platforms (12.4%) and professional/practice websites (11.9%), which operate outside traditional institutional structures.

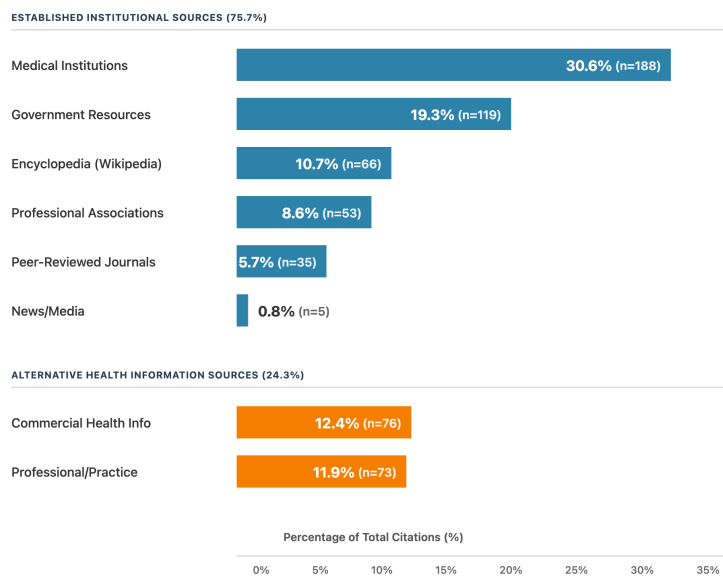

**Figure 2:** Institutional Dominance in ChatGPT health-related Citations: Distribution of 615 ChatGPT cited health sources by organization type showing distinction between established institutional sources and alternative health information sources. Percentages and source counts (n) are shown for each category.

Breaking down established institutional sources further, medical institutions led at 30.6%, followed by government resources (19.3%), encyclopedias (10.7%), professional associations (8.6%), peer-reviewed journals (5.7%), and news/media representing the smallest percentage (0.8%). Overall, citations showed substantial concentration among specific organizations. There were 10 organizations that accounted for 52.8% of all health citations with the highest concentrations among Wikipedia at 10.7%, Mayo Clinic at 9.9%, and Cleveland Clinic at 9.8% (Figure 3).

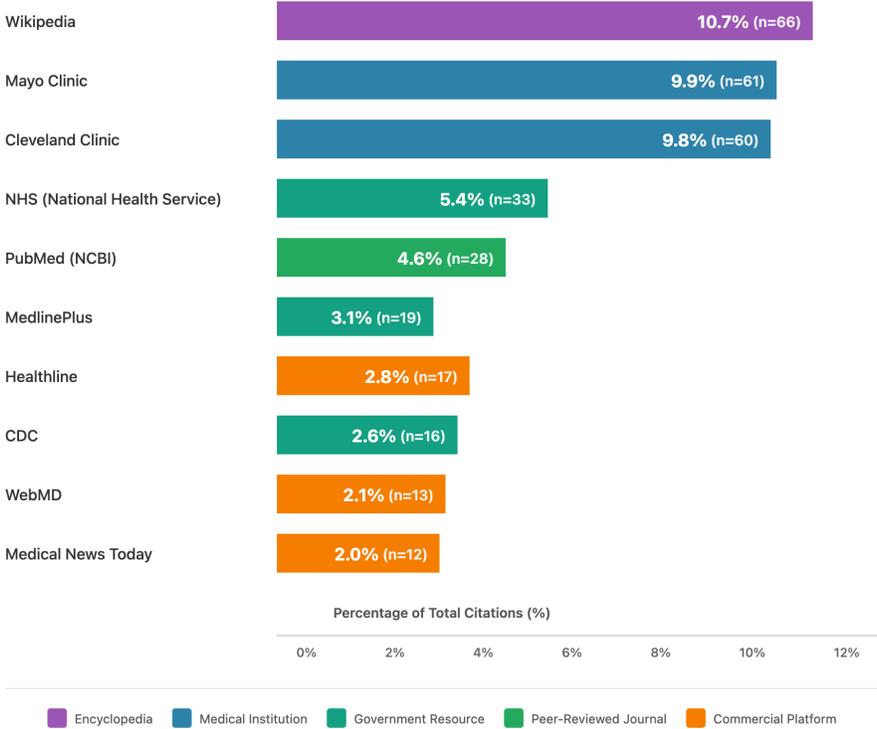

**Figure 3: Most Frequently Cited Health Information Sources.** *Top 10 organizations represent 325 of 615 total citations (52.8%) by ChatGPT.* Color coding indicates organization type: encyclopedia (purple), medical institutions (blue), government resources (teal), peer-reviewed journals (green), and commercial platforms (orange).

### 4.1 Examining authority signals across four domains shows clear patterns.

**Author Credentials Domain**: Author attribution was absent across the majority of sources (64.7%), while 28.3% provided partial attribution, and only 7.0% showed complete author attribution.

**Quality Assurance Domain**: Quality assurance mechanisms were inconsistently implemented across cited sources. Only 39.2% of sources listed references to support their health content, resulting in the majority having no cited references (60.8%). Additionally, 70.4% of sources did not state any medical review statements, nor did many have updated content, as 40.7% of the sources were either dated before 2020 or had no visible date; the remaining 22.9% were dated between 2020 and 2023, and 36.4% were published between 2024 and 2026. A separate analysis using Originality.ai was conducted to detect AI generated content. The tool successfully classified approximately 50% of the sources (n=305). Of the 305 classified sources, 22.6% were deemed to be AI-derived and 77.4% were human-derived.

**Digital Authority Domain**: Sources cited by ChatGPT had median page authority scores of 55, domain authority scores of 89, and low spam scores of 2. Comprehensive content length was viable in 47.3% of the sample. The remaining percentage was led by moderate comprehensive content at 33.7%, followed by brief content at 18.5%. The majority of sources implemented schema markup and JSON-LD format at a rate of 74.3%, and 68.3% respectively.

### 4.2 Authority Signal Profiles by Organization Type

Authority signal implementation among four categorical signal markers, references cited, schema markup, content length, and content recency varied substantially across organization types. Medical review statements were most prevalent among commercial health information platforms (71.1%) and medical institutions (49.5%), while government resources (13.4%), peer-reviewed journals (11.4%), professional/practice websites (4.1%), and encyclopedias (1.5%) were in the minority.

In examining references cited, encyclopedias (100.0%) and peer-reviewed journals (80.0%) heavily provided source citations compared to medical institutions (13.3%) and professional/practice websites (19.2%).

Schema markup presence was consistent across the various organization types, with encyclopedias (100.0%) and news/media outlets (100.0%) followed by commercial platforms (86.8%) and professional/practice websites (78.1%) leading, however peer-reviewed journals had comparatively lower rates of schema markup presence (42.9%).

Content length of web page text shows that encyclopedias (83.3%) and commercial health platforms (68.4%) routinely featured comprehensive content (>1,500 words). Professional/practice websites' comprehensive content was at 50.7%, and the lowest focus on comprehensive content were medical institutions (31.9%) and government resources (31.1%), having a more balanced approach to brief, moderate, and comprehensive content.

The recency of the content measured by date listed on the webpage showed that professional/practice websites had the greatest proportion of recent or updated dates (61.6% from 2024-2026), followed by government resources (50.4%) and medical institutions (41.0%). Commercial platforms (18.4%) and peer-reviewed journals (2.9%) had lower recency date rates.

## 5. Discussion

### 5.1 Three Credibility Strategies in AI-Cited Health Sources

Over 75% of the sources cited by ChatGPT originated from established institutional sources (medical institutions, government agencies, peer-reviewed journals, professional associations, news outlets, and Wikipedia). The remaining sources had various digital authority markers that may have led to them being sourced by ChatGPT.

These findings reveal three distinct credibility strategies (Figure 4) that may offer organizations competitive differentiation. First, established institutional sources rely on inherent authority rather than the explicit signals that alternative health information sources employ. Second, commercial platforms place emphasis on quality markers across multiple domains. Third, professional/practice websites maintain higher rates of content recency.

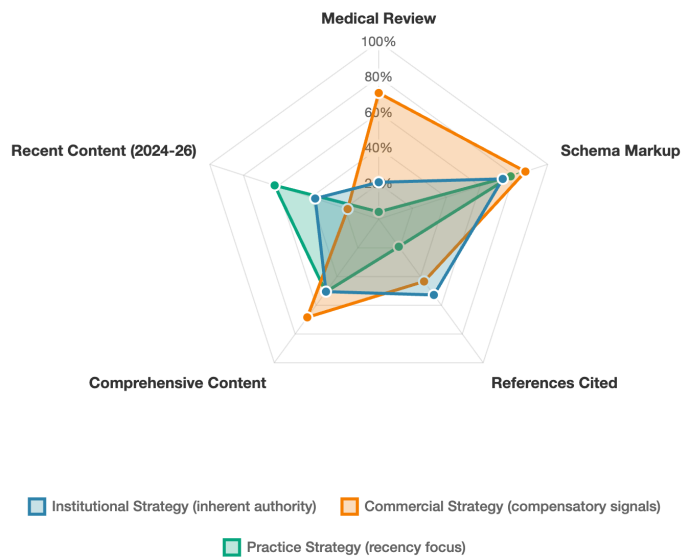

**Figure 4: Three Credibility Strategies in AI-Cited Health Sources.** This shows the differentiators among authority signals with established institutional sources in blue, commercial platforms in orange, and professional/practice websites in green.

First, institutional sources, including government agencies, medical institutions, and peer-reviewed journals, all benefit from inherent credibility advantages, as they operate under pre-existing, standardized vetting mechanisms (Kington et al., 2021). This institutional authority appears to reduce the perceived need for explicit quality assurance signals with government resources stating medical review in only 13.4% of cases, and similarly peer-reviewed journals in 11.4%, and these same organization types implemented schema markup at rates below the sample median (61.3% and 42.9% respectively).

Over half of the sources have DA scores above 88, yet that's not the mechanism by which commercial and practice professional websites have been strong. Instead, there is a mix of other authority signals that may be influencing their visibility in ChatGPT. Commercial health information platforms, which lack institutional backing, demonstrated investment in compensatory credibility signals across multiple domains. Commercial platforms stated medical review in 71.1% of sources, implemented schema markup in 86.8%, and produced comprehensive content (>1,500 words) in 68.4% of cases, which are substantially higher rates than institutional sources across all three signals. In contrast, professional practice websites have a higher percentage of content (61.6%) dated within the last two years, represented as recency dates. These findings are supported by other researchers who ascertained that LLM prefers newer dated content (Fang et al., 2025).

Among the commercial and professional practice websites, the deployment of technical optimization in the form of schema markup, quality assurance disclosures (medical review statements), content depth and recent dates is a measure of good practices, yet this can also be used as a deliberate strategy to compensate for the absence of institutional credibility in a way that may influence human perception and AI algorithmic retrieval selection.

The Authority Signals Framework examines ChatGPT source citations starting with examining 'who published it' through 'how does AI find it'. The results reveal organization types that fall within established health institutions dominant with factors such as authorship, editorial processes, and technical optimization serving as secondary signals. These results have implications for how health organizations and professionals proceed in what is inevitably the age of AI search optimization visibility.

## 5.2 Implications for Health Information Quality and Source Monitoring

Amid this growing use, OpenAI has announced ChatGPT Health, though currently being rolled out via a waitlist. It would allow users to enter their medical information for more tailored medical guidance (OpenAI, 2026). Given the prevalence of AI health information seeking, this study establishes baseline citation patterns to be monitored longitudinally. The authority signal patterns documented here have direct implications for patient safety. Health misinformation can have serious consequences including delayed care and adoption of ineffective and dangerous treatment (Memon et al., 2022; Tong et al., 2023). Established institutional health sources and health professionals have a vested interest in understanding AI citation protocols to avoid ceding visibility to commercial entities deploying compensatory signals. There is a need to continue to examine ChatGPT citation patterns, as AI search optimization strategies commonly referred to as AEO, GEO, GSO, AI SEO and other variants.

## 6. Limitations

This study is limited by the nature of the cross-sectional design which resulted in all of the data extraction occurring at one point at time, which prevents understanding of how this data will change over time. Given that a small subset of the HealthSearchQA dataset was used to examine these citation patterns, it's not clear whether these results may have remained consistent using a different subset of the data set questions or an entirely different set of questions.

## 7. Conclusion

This study introduces an Authority Signals Framework, organized in four domains that reflect key components to health information seeking, starting with Who wrote it (Author Credentials), followed by Who published it? (Institutional Affiliation), How was it vetted? (Quality Assurance), and How does AI find it? (Digital Authority). The primary sources cited among ChatGPT's health generated responses were from established institutional sources. The secondary determinant resulting in less than a quarter of the citations came from organization types that are characterized as commercial and professional practices web pages. This ChatGPT analysis establishes current baseline patterns on AI Search Optimization Strategies before widespread optimization strategies are implemented among health information organizations that lack established institutional backing. This investigation into AI health information citation behavior provides a reference for AI Search Optimization Strategies at the outset of 2026 prior to its gamification and enables longitudinal monitoring and comparison as its tactics evolve.

**Data Availability**

All research study materials and coding pipeline are publicly available at: https://doi.org/10.5281/zenodo.18287499

**AI Statement**

AI was used to support data collection, literature review search, overall formatting, image and table development along with editing for grammar and sentence mechanics..

**Software and Tools**

Moz. (2024). *Moz Domain Authority*. Moz Pro. Retrieved from https://moz.com/learn/seo/domain-authority
Originality.AI. (2024). *AI content detection tool* (Version 2.0) [Software]. Retrieved from https://originality.a
Anthropic. (2024). *Claude API: Claude Sonnet 4* [Application Programming Interface]. Retrieved from https://www.anthropic.com/api
Apify Technologies. (2024). *Website Content Crawler* [Web scraping tool]. Retrieved from https://apify.com/apify/website-content-crawler
Google. (2024). *Google Sheets API* (V4) [Application Programming Interface]. https://developers.google.com/sheets/api
Supabase Inc. (2024). *Supabase: The open source Firebase alternative* [Software]. Retrieved from https://supabase.com

**References**

American Medical Association. (2024). *Physician adoption of AI in clinical practice*. https://www.ama-assn.org

American Medical Association. (2024). *AMA augmented intelligence research: Physician sentiments around the use of AI in healthcare*. Retrieved from https://www.ama-assn.org/practice-management/digital-health/2-3-physicians-are-using-health-ai-78-2023

Anthropic. (2024). *Claude API: Claude Sonnet 4* [Application programming interface]. https://www.anthropic.com/api

Apify Technologies. (2024). *Website content crawler* [Web scraping tool]. https://apify.com/apify/website-content-crawler

Cai, Z. R., et al. (2024). Evaluation of ChatGPT-generated medical responses: A systematic review and meta-analysis. *Journal of Biomedical Informatics*, *150*, 104588.

Canadian Agency for Drugs and Technologies in Health (CADTH). (2025). *2025 Watch List: Artificial Intelligence in Health Care*. NCBI Bookshelf. https://www.ncbi.nlm.nih.gov/books/NBK613808/

Casselman-Hontalas, K., et al. (2024). Shifting trust in health information sources during the COVID-19 pandemic. *Journal of Health Communication*.

Chen, J., et al. (2024). Generative engine optimization and AI visibility in health information systems. *Digital Health Journal*.

Diviani, N., van den Putte, B., Giani, S., & van Weert, J. C. M. (2015). Low health literacy and evaluation of online health information: A systematic review of the literature. *Journal of Medical Internet Research*, *17*(5), e112. https://doi.org/10.2196/jmir.4018

Fahy, E., et al. (2014). Credibility signals in health information: A systematic review. *Health Information & Libraries Journal*.

Fang, X., Wang, H., & Sakai, T. (2025). Do large language models favor recent content? A study on recency bias in LLM-based reranking. *Waseda University Research Publications*.

Gao, L., et al. (2023). Citation transparency in large language model systems. *Nature Machine Intelligence*.

Ge, J., et al. (2025). Evaluating the reliability of ChatGPT for health-related questions: A systematic review. *Information*, *12*(1), 9.

Google. (2024). *Google Sheets API* (Version 4) [Application programming interface]. https://developers.google.com/sheets/api

Haltaufderheide, J., & Ranisch, R. (2024). Ethical concerns regarding validity and reliability of medical information from ChatGPT. *Journal of Medical Ethics*.

Hosseini, M., Gao, C. A., et al. (2023). High rates of fabricated and inaccurate references in ChatGPT-generated medical content. *Cureus*.

Iqbal, S., et al. (2025). Clinical decision-making and diagnosis using ChatGPT in medical settings. *Journal of Clinical Medicine*.

Jacques, E. (2026). Authority Signals in AI-Cited Health Information Sources: Automated Research Pipeline (Version 1.0.0) [Computer software]. Zenodo. https://doi.org/10.5281/zenodo.18287499

Kington, R., Arnesen, S., Chou, W. S., Curry, S. J., Lazer, D., & Villarruel, A. (2021). *Identifying credible sources of health information in social media: Principles and attributes*. NAM Perspectives. Discussion Paper, National Academy of Medicine. https://doi.org/10.31478/202107b

Kutner, M., et al. (2006). *The health literacy of America's adults: Results from the 2003 National Assessment of Adult Literacy*. U.S. Department of Education, National Center for Education Statistics.

MedlinePlus. (2024). *Evaluating health information*. U.S. National Library of Medicine. https://medlineplus.gov/evaluatinghealthinformation.html

Memon, S. A., et al. (2022). Health misinformation consequences: Delayed care and adoption of ineffective treatments. *BMJ Global Health*.

Moz. (2024). *Moz domain authority*. Moz Pro. https://moz.com/learn/seo/domain-authority

OpenAI. (2025). *ChatGPT usage statistics and health information seeking*. https://openai.com

OpenAI. (2026). *Introducing ChatGPT Health*. https://openai.com/index/introducing-chatgpt-health/

Originality.AI. (2024). *AI content detection tool* (Version 2.0) [Software]. https://originality.ai

Perlis, R. H., et al. (2024). Declining trust in physicians and hospitals: 2020-2024 trends. *JAMA Network Open*.

Search Engine Journal. (2025). *ChatGPT real-time web search capabilities*. https://www.searchenginejournal.com

Shahsavar, Y., & Choudhury, A. (2023). User confidence in ChatGPT for self-diagnosis and medical decision-making. *Digital Health*.

Silberg, W. M., et al. (1997). Assessing, controlling, and assuring the quality of medical information on the Internet. *JAMA*, *277*(15), 1244-1245. https://doi.org/10.1001/jama.1997.03540390074039

Sun, Y., Zhang, Y., Gwizdka, J., & Trace, C. B. (2019). Consumer evaluation of the quality of online health information: Systematic literature review of relevant criteria and indicators. *Journal of Medical Internet Research*, *21*(5), e12522. https://doi.org/10.2196/12522

Supabase Inc. (2024). *Supabase: The open source Firebase alternative* [Software]. https://supabase.com

Tangsrivimol, J., et al. (2025). Misleading information and hallucinations from AI chatbots in medical contexts. *Health Informatics Journal*.

Tong, A., et al. (2023). Health misinformation and adoption of ineffective treatments: A systematic review. *Preventive Medicine*.

Wu, T., et al. (2025). Alignment of LLM-generated health responses with cited sources. *Nature Communications*.

**Table 1**

*Authority Signal Characteristics of AI-Cited Health Sources (N=615)*

| Authority Signal | n (%) or Median (IQR) |
|---|---|
| **Institutional Affiliation Domain** | |
| Medical Institutions | 188 (30.6%) |
| Government Resources | 119 (19.3%) |
| Commercial Health Platforms | 76 (12.4%) |
| Professional/Practice Websites | 73 (11.9%) |
| Encyclopedia Sources | 66 (10.7%) |
| Professional Associations | 53 (8.6%) |
| Peer-Reviewed Journals | 35 (5.7%) |
| News/Media | 5 (0.8%) |
| **Author Credentials Domain** | |
| No Attribution | 398 (64.7%) |
| Basic Attribution (name only) | 174 (28.3%) |
| Complete Attribution (name + credentials) | 43 (7.0%) |
| **Quality Assurance Domain** | |
| Medical Review Stated | 182 (29.6%) |
| References Cited | 241 (39.2%) |
| Content Recency - 2024-2026 (Recent) | 224 (36.4%) |
| Content Recency - 2020-2023 | 141 (22.9%) |
| Content Recency - Before 2020/No Date | 250 (40.7%) |
| **Digital Authority Domain** | |
| Page Authority | Median: 55 (IQR: 40-67) |
| Domain Authority | Median: 89 (IQR: 80-95) |
| Spam Score | Median: 2 (IQR: 1-4) |
| Schema Markup Present | 457 (74.3%) |
| Content Length - Brief (<500 words) | 114 (18.5%) |
| Content Length - Moderate (500-1500 words) | 207 (33.7%) |
| Content Length - Comprehensive (>1500 words) | 294 (47.8%) |

**Note:** Authority signals organized by domain, prioritizing traditional health information quality markers (institutional affiliation, author credentials, quality assurance) before digital/technical metrics. Page Authority and Domain Authority measured via Moz API (0-100 scale). Spam Score ranges 0-100 (higher = more spam signals). IQR = Interquartile Range.

# Appendix A

*ChatGPT-Cited Sources: Organization Type Classification*

**DA Classification Threshold:** Professional/Practice Website: DA ≤ 45 (median DA = 21) | Commercial Health Information: DA > 45 (median DA = 73)

**Organization Type 1: Medical Institution**

**Definition:** Teaching hospitals, university-affiliated medical centers, hospital systems, medical schools, and integrated health systems that provide direct patient care and are typically affiliated with academic institutions.

**Classification Basis:** Institutional affiliation and organization name

**DA Score Range:** 41-93 | Median: 67

*Note: DA scores are descriptive characteristics of sources in this category*

**Sources (n=46) | Total Citations: 188**

| Organization Name | DA Score | Citations |
|---|---|---|
| Mayo Clinic | 92 | 61 |
| Cleveland Clinic | 88 | 60 |
| Johns Hopkins Medicine | 88 | 8 |
| University of Utah Health | 89 | 5 |
| Yale Medicine | 68 | 4 |
| Harvard Health Publishing | 93 | 3 |
| Stanford Medicine Children's Health | 67 | 2 |
| University of Pittsburgh Medical Center | 83 | 2 |
| Ann & Robert H. Lurie Children's Hospital of Chicago | 64 | 2 |
| Stanford Health Care | 74 | 2 |
| National Jewish Health | 65 | 2 |
| Kaiser Permanente | 84 | 2 |
| Houston Methodist | 68 | 2 |
| Institute for Health Metrics and Evaluation | 70 | 1 |
| OncoLink | 60 | 1 |
| Banner Health | 64 | 1 |
| Crossroads Hospice & Palliative Care | 49 | 1 |
| NYU Langone Health | 79 | 1 |
| Aurora Health Care | 61 | 1 |
| The Ohio State University Wexner Medical Center | 89 | 1 |
| CARE Hospitals | 48 | 1 |
| UCLA Health | 73 | 1 |

| Organization Name | DA Score | Citations |
|---|---|---|
| CityMD | 55 | 1 |
| University of Iowa Health Care | 65 | 1 |
| Mount Sinai Health System | 71 | 1 |
| HealthPartners | 63 | 1 |
| UC San Diego Health | 90 | 1 |
| University of Chicago Medicine | 73 | 1 |
| Columbia University | 92 | 1 |
| Apollo Hospitals | 58 | 1 |
| Vinmec International Hospital | 49 | 1 |
| Children's Health | 62 | 1 |
| Southern New Hampshire University | 64 | 1 |
| Hôpital de La Tour | 41 | 1 |
| Memorial Sloan Kettering Cancer Center | 78 | 1 |
| Mass General Brigham | 69 | 1 |
| St. Louis Children's Hospital | 61 | 1 |
| NewYork-Presbyterian | 69 | 1 |
| Penn Dental Medicine | 45 | 1 |
| Salus University Health | 42 | 1 |
| UT Health | 67 | 1 |
| Northwestern Medicine | 70 | 1 |
| Nebraska Medicine | 62 | 1 |
| Hospital for Special Surgery | 66 | 1 |
| Nationwide Children's Hospital | 66 | 1 |
| Dana-Farber Cancer Institute (Harvard Medical School) | 73 | 1 |

**Organization Type 2: Government Resource**

**Definition:** Official government health agencies at federal, state, and local levels, including .gov domains and international governmental health organizations (e.g., WHO). These are publicly funded entities with government accountability.

**Classification Basis:** Domain extension (.gov) and institutional affiliation

**DA Score Range:** 39-95 | Median: 89

*Note: DA scores are descriptive characteristics; government sources typically have high DA due to institutional authority*

**Sources (n=25) | Total Citations: 119**

| Organization Name | DA Score | Citations |
|---|---|---|
| National Health Service | 91 | 33 |
| MedlinePlus | 89 | 19 |
| Centers for Disease Control and Prevention | 94 | 16 |
| healthdirect Australia | 69 | 6 |
| World Health Organization | 95 | 5 |
| Better Health Channel Victoria | 72 | 5 |
| National Institute of Child Health and Human Development | 95 | 4 |
| National Institute of Diabetes and Digestive and Kidney Diseases | 95 | 3 |
| Northern Ireland Direct Government Services | 73 | 3 |
| National Institute of Mental Health | 95 | 3 |
| National Library of Medicine | 89 | 3 |
| Health Service Executive | 78 | 2 |
| National Institute of Neurological Disorders and Stroke | 95 | 2 |
| NHS inform | 67 | 2 |
| National Institute on Deafness and Other Communication Disorders | 95 | 2 |
| MyHealth Alberta | 91 | 2 |
| Los Angeles County Department of Public Health | 84 | 1 |
| National Institute of Dental and Craniofacial Research | 95 | 1 |
| National Institute of Arthritis and Musculoskeletal and Skin Diseases | 95 | 1 |
| West Virginia Department of Health and Human Resources | 74 | 1 |
| East-Central District Health Department | 81 | 1 |
| ADAMHS Board of Cuyahoga County | 39 | 1 |
| Virginia Department of Health | 87 | 1 |
| United States Environmental Protection Agency | 92 | 1 |
| HealthLink | 62 | 1 |

**Organization Type 3: Commercial Health Information**

**Definition:** Consumer-facing health content platforms and commercial entities whose primary business model involves providing health information to the public. These are for-profit organizations focused on health content creation and distribution.

**Classification Basis:** Institutional affiliation and organization name; DA > 45 used as an additional classifier for ambiguous cases

**DA Score Range:** 52-94 | Median: 73

*Note: DA threshold (>45) was used to distinguish commercial platforms from solo practice websites in ambiguous cases*

**Sources (n=28) | Total Citations: 76**

| Organization Name | DA Score | Citations |
|---|---|---|
| Healthline | 91 | 17 |
| WebMD | 94 | 13 |
| Medical News Today (Owned by Healthline Media) | 92 | 12 |
| Medscape | 89 | 4 |
| Verywell Mind | 81 | 3 |
| MSD Manuals | 73 | 2 |
| AFC Urgent Care Denver Cherry Creek | 54 | 2 |
| Healthgrades | 73 | 2 |
| Verywell Health | 82 | 2 |
| Epocrates | 59 | 1 |
| All About Vision | 69 | 1 |
| Vicks | 60 | 1 |
| Benadryl | 52 | 1 |
| Tylenol | 60 | 1 |
| Merck & Co., Inc. | 78 | 1 |
| Cancer Therapy Advisor | 56 | 1 |
| Listerine | 66 | 1 |
| Colgate-Palmolive Company | 69 | 1 |
| Ada | 57 | 1 |
| MedicineNet | 87 | 1 |
| Health.com | 88 | 1 |
| AMBOSS | 54 | 1 |
| Eucerin | 55 | 1 |
| HealthCentral LLC | 83 | 1 |
| BetterHelp | 75 | 1 |
| Health Information Page | 94 | 1 |
| GoodRx | 74 | 1 |
| Eating Recovery Center | 60 | 1 |

**Organization Type 4: Professional/Practice Website**

**Definition:** Individual medical practices, solo practitioners, private clinics, and small healthcare practices (typically 1-5 practitioners). These are independent healthcare providers with their own branded websites focused on attracting and serving patients.

**Classification Basis:** Institutional affiliation and organization name; DA ≤ 45 used as an additional classifier for ambiguous cases

**DA Score Range:** 5-58 | Median: 21

*Note: DA threshold (≤45) was used to distinguish solo/small practices from commercial platforms in ambiguous cases; lower DA reflects smaller, more localized online presence*

**Sources (n=69) | Total Citations: 73**

| Organization Name | DA Score | Citations |
|---|---|---|
| Refocus Corporate | 33 | 4 |
| Rock River Dental | 23 | 2 |
| Pete Walker, M.A. Psychotherapy | 45 | 1 |
| OneSkin | 43 | 1 |
| Colon & Digestive Health Specialists – Eastside Endoscopy Center | 12 | 1 |
| AmblyoPlay | 25 | 1 |
| 150 Harley Street ENT | 16 | 1 |
| Aether Health | 10 | 1 |
| Open Lines | 26 | 1 |
| Gainesville Dermatology & Skin Surgery | 23 | 1 |
| Alamo Pediatric Eye Center | 5 | 1 |
| Remagin | 18 | 1 |
| Rendia | 39 | 1 |
| Korean Cosmetics | 13 | 1 |
| Parris Orthodontics | 17 | 1 |
| G Smile | 29 | 1 |
| Valley Orthopedic Institute | 16 | 1 |
| My Menopause Centre | 40 | 1 |
| Ability Central | 28 | 1 |
| Physio.co.uk | 44 | 1 |
| H&D Physical Therapy | 16 | 1 |
| Optometrists.org | 58 | 1 |
| HealyOS | 12 | 1 |
| Avalon Dental Group | 17 | 1 |
| Spring View Dental | 10 | 1 |
| Beyond Women's Care | 8 | 1 |
| Virginia Foot and Ankle Center | 15 | 1 |
| Center for Orthopaedic Surgery and Sports Medicine | 27 | 1 |

| Organization Name | DA Score | Citations |
|---|---|---|
| Dallas Associates in Dermatology | 22 | 1 |
| Dr Francis Hall | 11 | 1 |
| Carda Health | 26 | 1 |
| Harbor | 40 | 1 |
| My Kidney Stone | 23 | 1 |
| Today's Vision Bulverde | 12 | 1 |
| Maryland Eye Care Center | 18 | 1 |
| WE C Hope | 38 | 1 |
| HMC Centre | 20 | 1 |
| Dr. Aliabadi Practice | 40 | 1 |
| Top Gynaecologists | 18 | 1 |
| Clear Chemist | 36 | 1 |
| Advanced Disc Replacement Spinal Restoration Center | 19 | 1 |
| Strenge Spine | 10 | 1 |
| Voluson Club | 40 | 1 |
| AOTC Jax | 14 | 1 |
| First OC Dermatology | 22 | 1 |
| Inspiring Smiles | 19 | 1 |
| Mt. Vernon Center for Dentistry | 21 | 1 |
| Lincoln Dental Associates | 26 | 1 |
| Ripon Dental | 7 | 1 |
| Liv Endodontics | 12 | 1 |
| Tompkins Dental | 30 | 1 |
| West Bell Dental Care | 21 | 1 |
| Gloss Dental | 9 | 1 |
| Texas Braces | 17 | 1 |
| All In 1 Dental | 25 | 1 |
| iO Dentistry | 23 | 1 |
| Eye Physicians | 24 | 1 |
| First Choice Podiatry | 5 | 1 |
| The Centers for Advanced Orthopaedics: Maryland Orthopedic Specialists Division | 29 | 1 |
| Watauga Orthopaedics | 25 | 1 |
| Newtown Foot and Ankle Specialists | 21 | 1 |

| Organization Name | DA Score | Citations |
|---|---|---|
| Eye Doctors - Elgart Gordon & Associates | 19 | 1 |
| Urgentcare for feet | 12 | 1 |
| Femmy Clinic | 30 | 1 |
| DaSC | 16 | 1 |
| 7 Day Home Care | 40 | 1 |
| Spine.md | 26 | 1 |
| West Broward Eyecare Associates | 20 | 1 |
| Piedmont Pediatrics | 18 | 1 |

**Organization Type 5: Encyclopedia**

**Definition:** General reference encyclopedias and medical encyclopedias that provide comprehensive health information organized in an encyclopedia format. These are primarily user-contributed or crowd-sourced reference platforms.

**Classification Basis:** Institutional affiliation and encyclopedia format

**DA Score Range:** 97-97 | Median: 97

*Note: DA scores are descriptive characteristics; Wikipedia's high DA reflects its established web authority*

**Sources (n=1) | Total Citations: 66**

| Organization Name | DA Score | Citations |
|---|---|---|
| Wikipedia | 97 | 66 |

**Organization Type 6: Professional Association/Resource**

**Definition:** Membership organizations of healthcare professionals, medical specialty societies, patient advocacy organizations with professional governance, and disease-specific foundations. These are typically nonprofit organizations focused on advancing medical knowledge, professional standards, or patient advocacy within specific health domains.

**Classification Basis:** Institutional affiliation and organization name

**DA Score Range:** 5-91 | Median: 65

*Note: DA scores are descriptive characteristics; wide range reflects diversity from established national organizations (high DA) to smaller specialty groups (lower DA)*

**Sources (n=41) | Total Citations: 53**

| Organization Name | DA Score | Citations |
|---|---|---|
| American Psychiatric Association | 73 | 4 |
| American Academy of Dermatology | 80 | 3 |
| American Dental Association | 65 | 3 |
| American College of Obstetricians and Gynecologists | 76 | 3 |

| Organization Name | DA Score | Citations |
|---|---|---|
| American Academy of Family Physicians | 80 | 3 |
| DermNet NZ (New Zealand Dermatological Society) | 68 | 2 |
| Sleep Foundation | 75 | 1 |
| Asthma + Lung UK | 68 | 1 |
| American Stroke Association | 67 | 1 |
| Prevent Child Abuse America | 58 | 1 |
| American Academy of Orthopaedic Surgeons | 70 | 1 |
| American College of Emergency Physicians | 59 | 1 |
| NAMI | 85 | 1 |
| International Alliance of ALS/MND Associations | 47 | 1 |
| Plastic Surgery Utrecht | 5 | 1 |
| American College of Rheumatology | 65 | 1 |
| Bone Health and Osteoporosis Foundation | 65 | 1 |
| Royal College of Paediatrics and Child Health | 64 | 1 |
| BrainFacts.org | 63 | 1 |
| National Eating Disorders Association | 82 | 1 |
| Endocrine Society | 73 | 1 |
| NICABM | 52 | 1 |
| Scleroderma and Raynaud's UK | 51 | 1 |
| American Optometric Association | 58 | 1 |
| American Brain Foundation | 54 | 1 |
| LD Resources Foundation of America | 41 | 1 |
| National Eczema Society | 59 | 1 |
| Fertility Network UK | 56 | 1 |
| American Kidney Fund | 63 | 1 |
| Diabetes UK | 67 | 1 |
| Trial Net | 50 | 1 |
| National Kidney Foundation | 72 | 1 |
| Susan G. Komen | 75 | 1 |
| Sepsis Alliance | 60 | 1 |
| American Cancer Society | 91 | 1 |
| National Brain Tumor Society | 62 | 1 |
| Palliative Care Network of Wisconsin | 40 | 1 |

| Organization Name | DA Score | Citations |
|---|---|---|
| Radiological Society of North America (RSNA), American College of Radiology (ACR), American Society of Radiologic Technologists (ASRT) | 66 | 1 |
| Cancer Research UK | 85 | 1 |
| Target ALS | 43 | 1 |
| The Brain Tumor Charity | 61 | 1 |

**Organization Type 7: Peer-Reviewed Journal**

**Definition:** Academic journals, scientific publications, and peer-reviewed research databases that publish original research and systematic reviews. These sources include both individual journal articles and aggregator platforms that index peer-reviewed literature.

**Classification Basis:** Institutional affiliation and publication format

**DA Score Range:** 46-95 | Median: 84

*Note: DA scores are descriptive characteristics; high DA reflects academic authority and established indexing systems*

**Sources (n=8) | Total Citations: 36**

| Organization Name | DA Score | Citations |
|---|---|---|
| National Center for Biotechnology Information PubMed | 95 | 28 |
| BMJ Group | 76 | 2 |
| NANOS | 46 | 1 |
| ASM Journal | 75 | 1 |
| ScienceDirect | 93 | 1 |
| Journal of Clinical Investigation | 77 | 1 |
| Frontiers Media | 92 | 1 |
| Elsevier Ltd. | 93 | 1 |

**Organization Type 8: News/Media**

**Definition:** Traditional news organizations and media outlets that cover health topics as part of broader news coverage. These are journalism-focused organizations where health is one reporting beat among many topics.

**Classification Basis:** Institutional affiliation and organization name

**DA Score Range:** 78-94 | Median: 93

*Note: DA scores are descriptive characteristics; high DA reflects established media authority*

**Sources (n=5) | Total Citations: 5**

| Organization Name | DA Score | Citations |
|---|---|---|
| The Economic Times | 94 | 1 |

| Organization Name | DA Score | Citations |
|---|---|---|
| Times of India | 94 | 1 |
| New York Post | 93 | 1 |
| News-Medical.Net | 78 | 1 |
| Live Science | 91 | 1 |

**Summary Statistics**

**Total Sources Analyzed:** 223 unique organizations | **Total Citations:** 615